\documentclass[aps,prl,amsmath,amssymb,reprint, twocolumn,superscriptaddress, reprintnumbers,showpacs,intlimits,longbibliography]{revtex4-2}
\pdfoutput=1
\bibpunct{[}{]}{,}{n}{}{}

\usepackage[utf8]{inputenc}
\usepackage{bm,latexsym,mathrsfs,enumerate}
\usepackage[dvipsnames]{xcolor}
\usepackage{mathtools}
\usepackage{makecell}
\usepackage{multirow}
\usepackage[breaklinks=true,unicode=true,urlcolor = blue,colorlinks = true,citecolor = blue,linkcolor = blue]{hyperref}
\usepackage{graphicx}
\usepackage{ragged2e}
\usepackage[export]{adjustbox}
\renewcommand{\vec}[1]{\bm{#1}}
\usepackage[final]{pdfpages}
\usepackage{tikz}
\makeatletter
\AtBeginDocument{\let\LS@rot\@undefined}
\makeatother
\begin{document}

\title{Temperature evolution of the Fermi surface of the FeSe monolayer on SrTiO$_3$}


\author{ Khalil~Zakeri}
\email{khalil.zakeri@partner.kit.edu}
\affiliation{Heisenberg Spin-dynamics Group, Physikalisches Institut, Karlsruhe Institute of Technology, Wolfgang-Gaede-Strasse 1, D-76131 Karlsruhe, Germany}
\author{Ryan Roemer}
\affiliation{Department of Physics and Astronomy and Quantum Matter Institute, University of British Columbia,
	Vancouver, British Columbia, Canada V6T 1Z1}
\author{Ke Zou}
\affiliation{Department of Physics and Astronomy and Quantum Matter Institute, University of British Columbia,
	Vancouver, British Columbia, Canada V6T 1Z1}
\begin{abstract}
The origin of superconductivity in the FeSe monolayer on SrTiO$_3$ remains one of the unresolved mysteries in condensed-matter physics. Here by investigation of the temperature evolution of the dynamic charge response of FeSe/SrTiO$_3$  we infer that the response of the monolayer itself is nearly temperature independent. This indicates a constant Fermi surface over a wide range of temperature, in stark contrast to that of the bulk FeSe and other Fe-based superconductors. Our results, which manifest the peculiarity of the electronic structure of the FeSe monolayer, may help for a microscopic understanding of the superconductivity in Fe-chalcogenide monolayers on oxide surfaces in general.
\end{abstract}

\date{\today}
\maketitle

\section{Introduction}\label{Sec:Intro}

The observation of high-temperature superconductivity in Fe-based compounds challenged the theories of superconductivity  \cite{Kamihara2008,Hsu2008}.
Several ideas have been proposed, suggesting how one may revise, extend or totally rethink the existing theories \cite{Stewart2011,Johnson2015}. Among Fe-based superconductors FeSe has been the focus of many theoretical models, since it is the most simple compound in term of the crystal structure. The material has become even more attractive after the  discovery of  a much higher superconducting transition temperature $T_c$ in a monolayer (ML) of FeSe grown on Nb-doped SrTiO$_3$(001) (hereafter Nb-STO) \cite{Wang2012a,Liu2012,Tan2013,He2013,Bozovic2014,Ge2014,Lee2014} and some other oxide substrates \cite{Peng2014,Zhao2018,Liu2021a}.
The observed high $T_c$ has been argued to be an interfacial effect \cite{Zhang2017,Zhou2018,Song2019,Xu2020}.

The size and the nesting properties of the Fermi surface (FS) are important to understand the nature of superconductivity, pairing mechanism and the value of $T_c$. Despite many similarities, one of the distinct differences between FeSe ML and its bulk counterpart is the missing hole pockets at the $\bar{\Gamma}$--point \cite{He2013,Zhang2017,Zhang2017a,Song2019}. It has been proposed that a strong downward shift of the hole pockets in bulk FeSe  or in thick FeSe films would lead to an enhancement of $T_c$ \cite{Seo2016,Shi2017}. This shift changes the topology of FS and is known as the Lifshitz transition.
In the case of bulk FeSe it has been observed that both the hole and electron pockets exhibit an anomalous temperature dependence \cite{Kushnirenko2017,Pustovit2017,Rhodes2017,Pustovit2019}. The effect has been attributed to either (i)  the so-called Pomeranchuk effect, as a consequence of electronic instabilities  \cite{Zhai2009,Chubukov2016,Davis2013,Massat2016,Classen2017} or (ii) the renormalization of the bands by spin fluctuations \cite{Ortenzi2009,Benfatto2011}, as a result of self-energy effects \cite{Kushnirenko2017}, or (iii) be due to the suppression of the nearest-neighbor hopping, as a result of spin or orbital orderings \cite{Pustovit2016}. The temperature evolution of the electronic bands and that of FS of FeSe ML over a wide range of temperature would provide a way of checking the validity of  these hypotheses. It is also important in the contexts of interaction-induced instabilities within strongly correlated electron systems, altermagnetism and Bogoliubov FS \cite{Mazin2023,Wu2024}.

In this paper we report on the measurements of the dynamic charge response of the FeSe ML on Nb-STO as a function of temperature and provide the necessary experimental basis for verifying the validity of the theoretical approaches, which aim to address the electronic structure of FeSe ML and its bulk counterpart as well as the presence of possible electronic correlations in Fe-chalcogenides. We show that the temperature evolution of the dynamic charge response can be well explained when a constant FS is assumed over the whole range of temperature. This observation is in sharp contrast to the observed strong anomalous temperature dependence of the bulk FeSe FS. We will discuss the consequences of this temperature insensitive FS. Moreover, we will introduce the high-resolution electron scattering techniques, which probe the dynamic charge response, as a tool for probing changes of the FS in low-dimensional quantum materials, thin films and heterostructures.

\section{Results}\label{Sec:Results}

\begin{figure}[t!]
	\centering
	\includegraphics[width=1.0\columnwidth]{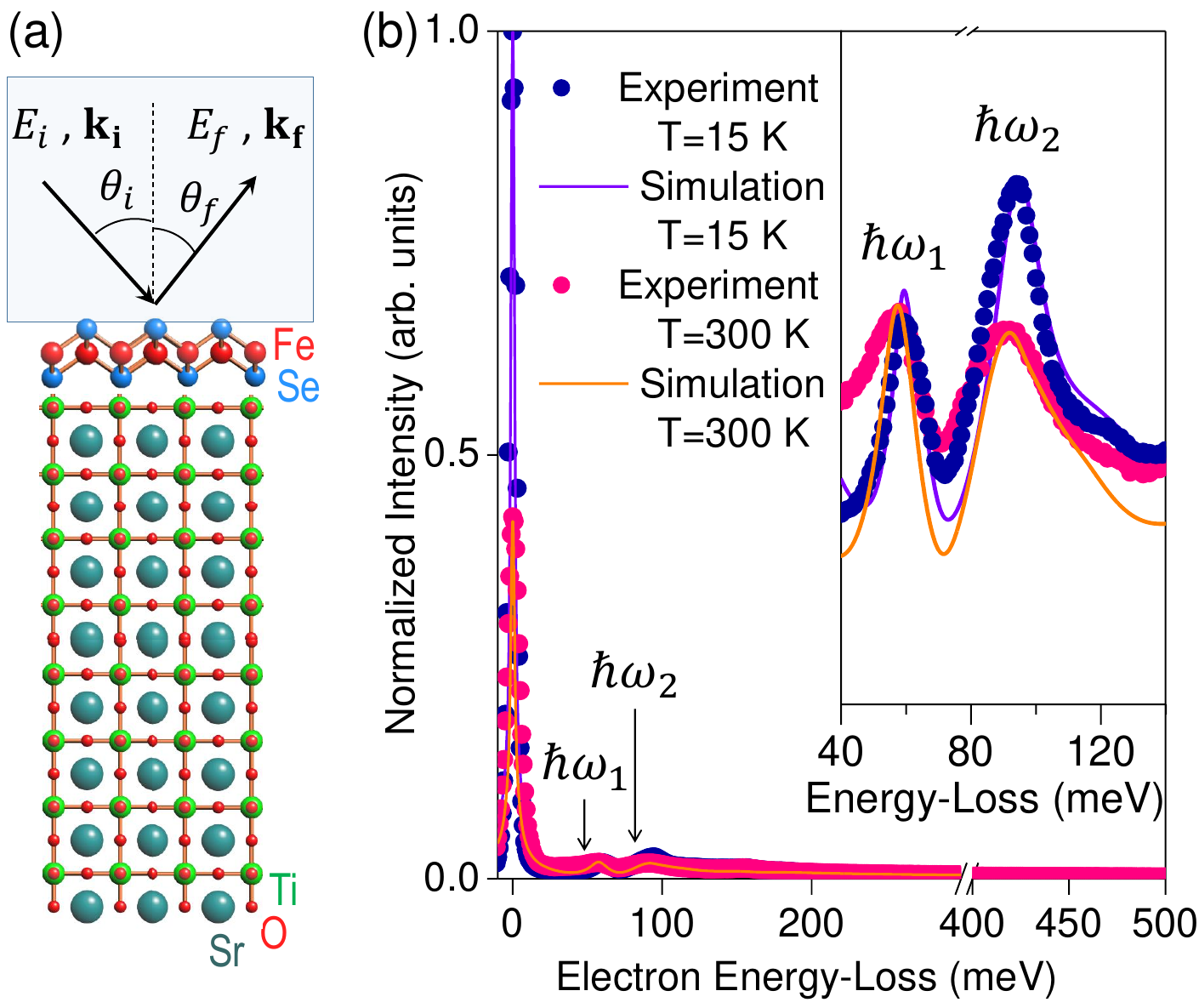}
	\caption{(a) The scattering geometry used for our HREELS experiments. The energy and momentum of the incident (scattered) beam are denoted by $E_i$ ($E_f$) and $\vec{k_i}$ ($\vec{k_f}$), respectively. (b) HREEL spectra recorded at $T=15$~K (dark-blue) and $T=300$~K (red). The simulated data are also shown for a comparison (see text for details). The spectra were recorded at $q=0$ \AA$^{-1}$ (the $\bar{\Gamma}$--point) and $E_i=4.08$~eV.  The inset shows a magnified part of the spectra.}
	\label{Fig:Fig1}
\end{figure}

Epitaxial FeSe MLs were grown by molecular beam epitaxy on Nb-doped STO(001) using the well established procedure, described in detail in Refs.~\cite{Liu2020,Liu2021,Jandke2019,Zakeri2023} and Note~I of Supplemental Material \cite{SuppMat}. The dynamic response of the system was probed by means of high-resolution electron energy-loss spectroscopy (HREELS), where an electron beam with an incident energy $E_i$ is scattered from the sample surface and the energy distribution of the scattered beam is probed with respect to the in plane electron momentum transfer $q= |\vec{q}|=k_{i \parallel}-k_{f \parallel}$ (here $k_{i\parallel}$ and $k_{f\parallel}$ denote the parallel components of the electron momentum before and after the scattering event, respectively). The scattering geometry is sketched in Fig.~\ref{Fig:Fig1}(a). The scattering plane was parallel to the STO(001)[100]-direction, corresponding to the $\bar{\Gamma}$--$\bar{\rm X}$ direction of the surface Brillouin zone. The spectra were recorded at $E_i\approx4.1$~eV and $q=0$.  Typical HREEL spectra recorded at $T=15$ and $300$~K on an FeSe ML grown on a substrate with 0.7\% Nb (SrTi$_{0.993}$Nb$_{0.007}$O$_3$) are presented in Fig.~\ref{Fig:Fig1}(b).
Besides the elastic peak, one identifies the signature of several phonon modes associated with the FeSe ML and the FeSe/STO interface. The weak modes located at 11.8, 20.5, 24.8 and 36.7~meV are similar to the phonon modes observed at the $\beta$--FeSe(001) surface \cite{Zakeri2017,Zakeri2018} and can therefore be assigned to the FeSe ML phonons.  These modes can be better resolved at the zone boundary \cite{Zhang2016,Zhang2018,Zakeri2023}. The two prominent interfacial Fuchs-Kliewer (FK) modes are located at the energies of $\hbar\omega_1=59.1$~meV (56.9~meV) and $\hbar\omega_2=94.2$~meV (91.1~meV) at $T=15$~K (300~K) in agreement with the previous HREELS experiments \cite{Zhang2016,Zhang2018,Zhang2019,Xu2022}.
The most important observation is that the relative intensity of these modes   depends rather strongly on temperature. It is therefore of prime importance to unravel the origin of this observation.

Generally the scattering intensity in the HREELS experiments provides direct access to the spectral function $\mathcal{S}(\vec{q}, \omega)$ that is given by   $\int_{-\infty}^0\,\mathcal{S}(\vec{q},z,z',\omega)e^{-q|z+z'|}dzdz'$, where the integration is over the direction perpendicular to the surface, represented by variables $z$ and $z'$ \cite{Evans1972,Vig2017,Zakeri2022}. $\mathcal{S}(\vec{q}, \omega)$  is directly related to the dynamic charge response of the sample, since $\mathcal{S}(\vec{q},z,z',\omega)$ denotes the density-density correlation function and is given by $\langle m|\hat{\rho}(-\vec{q},z)|n\rangle\langle n|\hat{\rho}(\vec{q},z')|m\rangle\delta(\hbar\omega+E_m-E_n)$, where $\hat{\rho}$ indicates the charge density operator, $|m\rangle$ ($|n\rangle$) denotes the many-body states of the sample with the energy $E_m$ ($E_n$) and $\vec{q}$ is the two-dimensional momentum vector parallel to the surface. It is important to emphasize that $\mathcal{S}(\vec{q}, \omega)$ represents the total charge response of the sample and is therefore a measure of the dynamic response of all charges inside the sample, e.g., electrons, holes, and ions.
Hence, a careful analysis of the HREELS intensity would provide direct information on the dynamic charge response and consequently on the charge distribution inside the system. In particular, the change in the relative intensities of the FK modes, as seen in Fig.~\ref{Fig:Fig1}, must therefore be due to changes  in the dynamical response of the system.

\begin{figure}[t!]
	\centering
	\includegraphics[width=1.0\columnwidth]{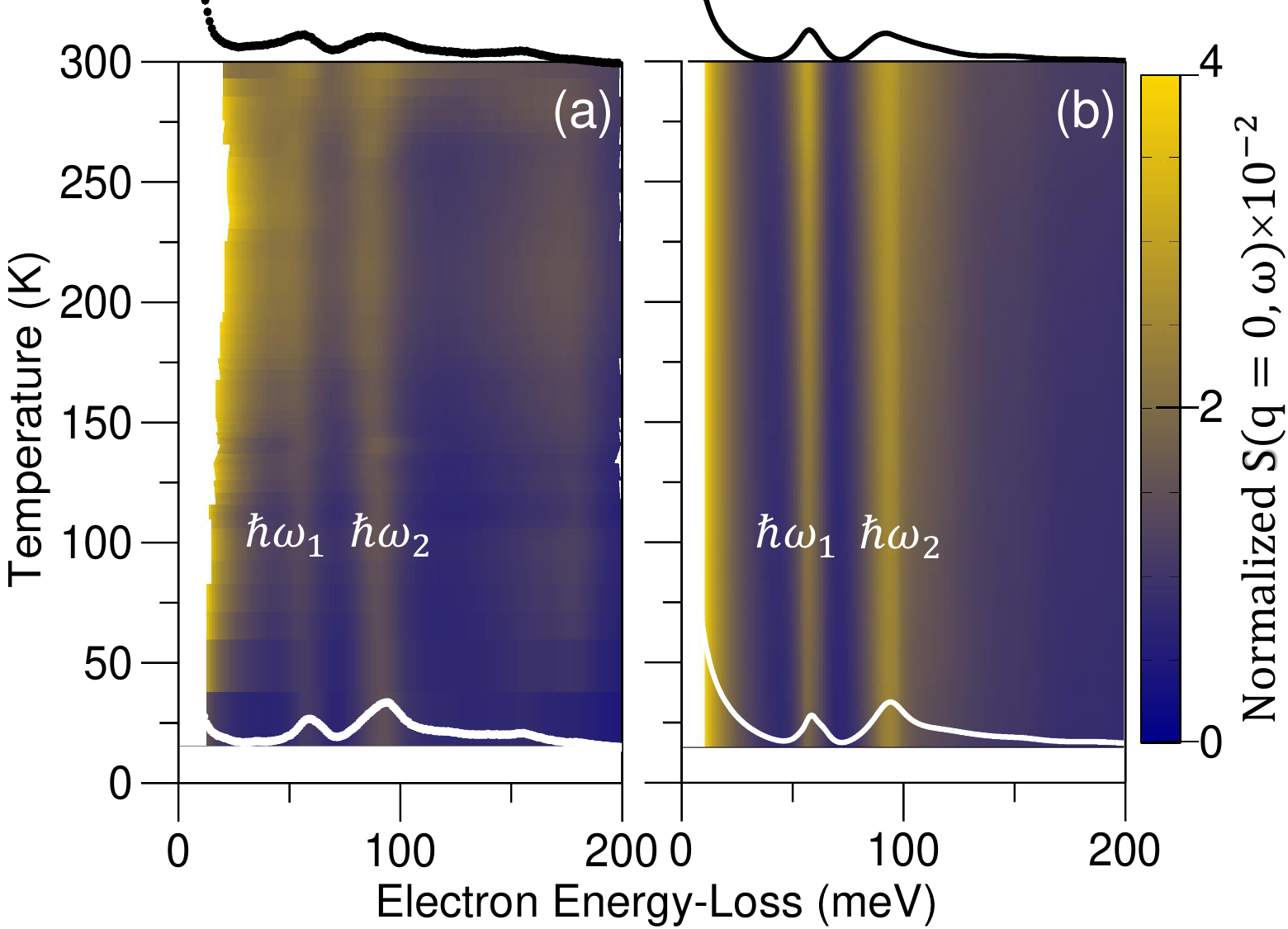}
	\caption{The experimental (a) and simulated (b) structural factor as a function of temperature. The data were recorded on an FeSe ML grown on Nb-STO(001) with a doping level of 0.7\%, using a beam energy of $E_i =$ 4.08~eV. The spectra for two limiting cases are shown on the top and bottom.}
	\label{Fig2}
\end{figure}

To understand the temperature dependence of the dynamic charge response, we performed numerical simulations of the spectra, based on the formalism developed in Refs.~\cite{Lambin1990,Sunjic1971,Lucas1972, Zakeri2021} (see Note~I of Supplemental Material for details \cite{SuppMat}). The most important input parameters are those associated with the charge carriers inside the STO substrate as well as those associated with the electron density in the FeSe film. Both appear in terms of plasmon frequencies in the formalism \cite{SuppMat}. It is known that the plasmon frequency associated with the charge carriers inside the STO is strongly temperature dependent \cite{Gervais1993,Eagles1996,Galzerani1982,Collignon2020}. The effect has been explained on the basis of the mixed-polaron theory \cite{Eagles1996}. For the simulation at $T=15$~K we estimate the bulk plasmon frequency based on the measurements by means of optical techniques \cite{Gervais1993,Eagles1996,Collignon2020}. For simulations at other temperatures we leave the bulk plasmon frequency of STO as a variable. The plasmon frequency of  charge carriers inside FeSe ML can  be estimated using the photoemission data. Owing to its simple FS, the plasmon frequency is directly proportional to the Fermi wavevector $k_{\rm{F}}$
via $\omega_{pl}=\lambda\frac{ k_{\rm{F}}}{\sqrt{m_{\rm{eff}}}}$, where the constant $\lambda$ is related to the elementary charge $e$, the dielectric constant of FeSe $\varepsilon^{\rm{FeSe}}_\infty$,  the vacuum permittivity $\epsilon_0$, and the film thickness $d$ via  $\lambda=e(\varepsilon^{\rm{FeSe}}_\infty\epsilon_0 \pi d)^{-1/2}$. $m_{\rm{eff}}$ is the effective mass. The values of $m_{\rm{eff}}$ and $k_{\rm{F}}$ determine the size of the FS and are known experimentally \cite{Lee2014,Wang2016,Zhang2017,Zhou2017,Faeth2021,Rademaker2021,Liu2021,Li2018}. Obviously, any considerable change in the FS would directly influence the dynamic charge response measured by HREELS through changes in $\omega_{pl}$. In the simulation of both spectra shown in Fig.~\ref{Fig:Fig1}(b) we use the same values of  $k_{\rm{F}}=0.22$~\AA$^{-1}$ and $m_{\rm{eff}}=3m_e$, where $m_e$ represents the free electron's mass. The geometrical structure used for the numerical simulations is a slab composed of an FeSe ML on an insulating STO depletion layer on a semi-infinite Nb-STO(001). The presence of the depletion layer, the necessity of its consideration, and its impact have been discussed in Refs.~\cite{Zakeri2023,Zakeri2023a}. Comparing the results to those measured experimentally indicates that the intensity ratio of the FK modes can be well explained solely based on the temperature dependence of the plasmon frequency associated with the charge carriers inside the STO substrate and that of the STO phonons. Another important result of this observation is that the intensity ratio of the FK modes can be used as a benchmark of changes in the plasmon frequency and, consequently, the size of the  FS. Aiming at a detailed understanding of the temperature dependence of FS we investigated the intensity ratio of the FK modes in detail. Figure \ref{Fig2}(a) shows the experimental dynamic charge response of the FeSe/STO system probed over the temperature range of 15--300~K. The results of the simulation are also provided in Fig.~\ref{Fig2}(b).
In the simulation the only variable is the plasmon frequency associated with the bulk carriers in STO \cite{SuppMat}. In Fig.~\ref{Fig2} the main feature to be emphasized is the change in the peak ratio of the FK modes.
The good agreement between the experimental dynamic charge response and the results of the simulation provides an access to the temperature dependence of the plasmon frequency associated with the carriers inside STO. Similar to that of the experiment, at low temperatures the intensity of the first FK mode $I_{\hbar\omega_1}$ is smaller than that of the second FK mode $I_{\hbar\omega_2}$. The relative intensity of these two modes $I_{\hbar\omega_2}/I_{\hbar\omega_1}$  decreases with temperature.

\begin{figure}[t!]
	\centering
	\includegraphics[width=1.0\columnwidth]{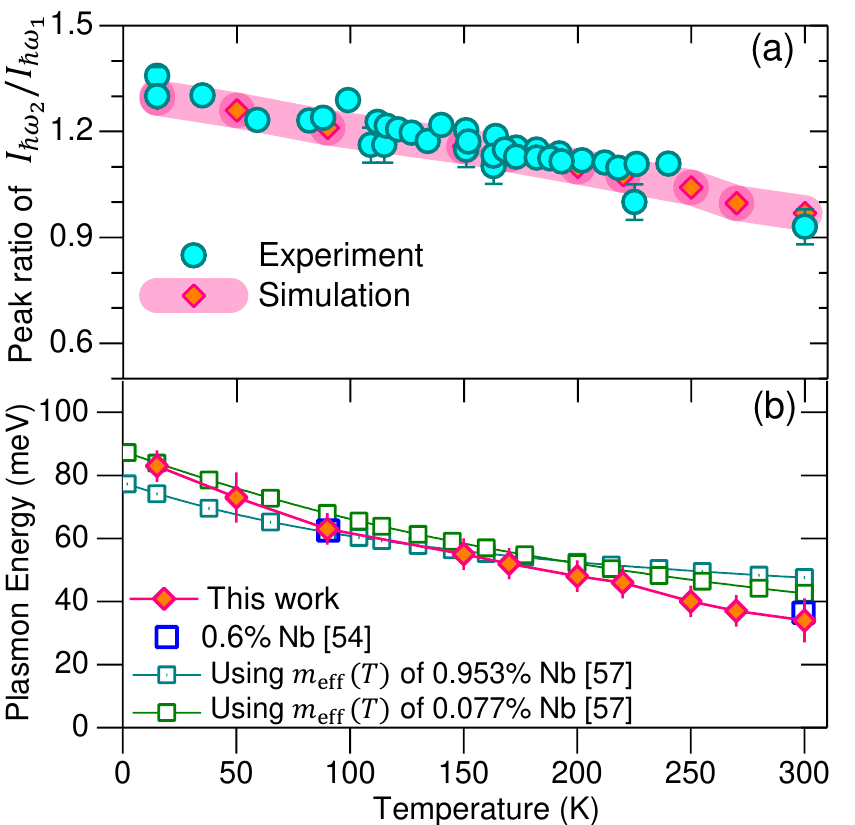}
	\caption{(a) The measured and simulated peak ratio of the two FK modes  $\mathcal{R}=I_{\hbar\omega_2}/I_{\hbar\omega_1}$ vs temperature. The Nb doping was 0.7\%. (b) Temperature dependence of plasmon energy of the charge carriers inside STO. The results of Refs.~\cite{Gervais1993,Collignon2020} obtained for STO are shown for a comparison. The green data were obtained based on $\omega_{p}(T) =\sqrt{n_c e^2/\varepsilon^{\rm{STO}}_{\infty} \epsilon_0 m_{\rm eff}(T)}$, using the temperature dependence of  $m_{\rm{eff}}$ reported in Ref.~\cite{Collignon2020} for two doping levels of STO and $n_c=1.18\times10^{26}$~m$^{-3}$, corresponding to 0.7\% Nb doping.}
	\label{Fig3}
\end{figure}

The intensity ratio of the simulated spectra defined as $\mathcal{R}=I_{\hbar\omega_2}/I_{\hbar\omega_1}$ vs temperature is plotted in Fig.~\ref{Fig3}(a), together with the experimental data. The values used for the plasmon frequency of STO carriers for different temperatures are plotted in Fig.~\ref{Fig3}(b). We emphasize again that FS was kept constant over the entire temperature range. The results clearly demonstrate that the size of FS is nearly temperature independent.
In Fig.~\ref{Fig3}(b) we also show the values of the plasmon frequency probed by Gervais \textit{et al}., in the STO samples with a similar doping level \cite{Gervais1993}. The results indicate that if the temperature dependence of the plasmon frequency of the STO carriers is taken into account one can describe the HREELS data very well.

The plasmon frequency associated with the charge carriers in Nb-STO is given by $\omega_{p}(T) =\sqrt{n_c e^2/\varepsilon^{\rm{STO}}_{\infty} \epsilon_0 m_{\rm eff}(T)}$, where $n_c$ denotes the carrier concentration, $\varepsilon^{\rm{STO}}_{\infty}$ is the dielectric constant of STO. The strong temperature dependence of plasmon frequency in STO is due to the temperature dependence of $m_{\rm eff}$ of these carriers. In Fig.~\ref{Fig3}(b) we show, furthermore, the temperature dependence of the plasmon frequency of the STO bulk carriers using $n_c=1.18\times10^{26}$~m$^{-3}$, corresponding to 0.7\% Nb doping, and $m_{\rm{eff}}(T)$ reported by Collignon \textit{et al}.,  \cite{Collignon2020}.
The main conclusion of Fig.~\ref{Fig3} is that the temperature dependence of the dynamic charge response (and that of $\mathcal{R}$) can be well explained by considering the temperature dependence of the plasmon frequency inside STO and, at the same time, a temperature-independent FS. This means that any temperature-induced changes in the electronic structure, e.g., orbital-dependent correlations must happen below the Fermi level \cite{Liu2020a,Yi2015,Pu2016}. It is important to emphasize that $\mathcal{S}(\vec{q}, \omega)$ represents the dynamic response of the FeSe/STO system as a whole. Hence, contributions of all subsystems, e.g., FeSe ML, the depletion layer, and the volume part of doped STO, are decisive for the final response function. All these contributions have been properly taken into account in our formalism, as discussed above and in Note~I of Supplemental Material \cite{SuppMat}.

One way to suppress the contribution of the STO charge carriers to $\mathcal{R}$ is to perform experiments and simulations on an FeSe ML grown on a substrate with a low level of Nb doping.
In this case due to the low carrier concentration of STO one has to consider (i) a rather large depletion layer and (ii) a much weaker temperature dependence of plasmon frequency of the STO carriers. Hence, the temperature dependence of  the dynamic charge response and that of $\mathcal{R}$ would be largely determined by the temperature dependence of the STO phonons and the charge carriers inside the FeSe ML. The measurements and simulations performed on an FeSe ML on a lightly doped STO  (0.1\%) are provided in Note~II of Supplemental Material \cite{SuppMat}). The results indicate that the temperature dependence of the dynamic charge response can be well described if one considers the temperature dependence of phonon frequencies and oscillator strengths of STO and assumes a temperature-independent FS for FeSe ML.

As a side remark, in addition to their impact on $\mathcal{R}$ the signature of the STO charge carriers can also be seen in the spectral function. Considering the data presented in Fig.~\ref{Fig2}, one recognizes a
broad signal at about $\hbar\omega=110$~meV. This is a consequence of the charge carriers in FeSe ML as well as in the inner part of Nb-STO. In the spectra recorded at higher incident energies an additional peak has been
observed at $\approx$150 meV  \cite{Zhang2018,Xu2022}. Our simulations suggest that this peak originates from the carriers located below the depletion layer and should therefore be more pronounced in the spectra recorded at higher incident energies \cite{Note1}.

\section{Discussion}\label{Sec:Discussion}

The phenomenon of high-temperature superconductivity in Fe chalcogenides can be understood only if one can explain the properties of all the members of this family within the same model. Our observation of the constant FS of FeSe/STO is in sharp contrast to that of FeSe bulk \cite{Kushnirenko2017,Pustovit2017,Rhodes2017} and other Fe-based high-$T_c$ materials \cite{Brouet2013,Dhaka2013}, showing a significant temperature dependence. While in Ref.~\cite{Kushnirenko2017} an expansion of both hole and electron pockets is observed,  Ref.~\cite{Rhodes2017} reports on an expansion of the electron pockets and a shrinking of the hole pockets with temperature. Both studies report on a considerably large increase of the size of the electron pockets by a factor of 4.
The temperature dependence of FS has been calculated using a combined tight-binding and density functional approach \cite{Schrodi2018}.
No notable
temperature-induced changes in the electron pockets at the $\bar{\rm M}$-point has been observed, in line with our results and those of Ref.~\cite{Liu2020a}.
The strong temperature dependence of the FS of bulk Fe-based superconductors is argued to facilitate the nesting properties and also the formation of charge density waves \cite{Kushnirenko2017,Pustovit2017,Rhodes2017,Brouet2013, Dhaka2013}. This is particularly important for systems with  hole pockets at the zone center. The presence of electronic instabilities at the FS are also important in the context of altermagnetism, recently proposed for this and other similar compounds \cite{Mazin2023}.

The absence of hole pockets in the case of FeSe ML suggests that the electronic coupling is very likely between the electron pockets at the zone corner. Such a coupling does not therefore require a significant change of the FS with temperature. Only small changes in the FS can, eventually, lead to satisfying the conditions required for the Cooper pairing of electrons, in particular, if the pairing mechanism is of an electronic nature and of $S_{\pm}$ character \cite{Jandke2019,Liu2019}. In the case where the pairing mechanism is mediated by phonons such a nesting of FS is not crucial.

The size of FS is decisive for the determination of $T_c$, irrespective of the nature of superconducting pairing mechanism. It has been found that within the Eliashberg theory the maximum possible $T_c$ in two-dimensional superconductors is determined by an interplay between the size of FS and its nesting properties.  Systems with intermediate size of FS have been found to exhibit maximum $T_c$ \cite{Schrodi2021}. It seems that in the case of FeSe ML due to a massive charge transfer from the depletion region of the STO substrate into the FeSe ML the size of FS is optimized \cite{Zakeri2023}. This optimized size of FS remains nearly constant over a wide range of temperature. Note that the observed  temperature-insensitive FS is not in contrast to the orbital-dependent correlation effects reported for several Fe chalcogenides \cite{Yi2015,Pu2016}

\section{Conclusion}\label{Sec:Conclusion}

In conclusion, we probed the dynamic charge response of the FeSe ML on STO over a wide range of temperature. The intensity ratio of the interfacial FK modes was regarded as a benchmark for the investigation of any changes in the charge carriers by temperature. It was observed that the only electronic contribution which varies with temperature is the plasmon frequency associated with the charge carriers inside STO.
The results indicate that, unlike its bulk counterpart, FeSe ML exhibits a nearly temperature independent FS, meaning that any temperature-induced change of the chemical potential in FeSe ML is much smaller than that of its bulk counter part.
The size of FS is decisive for several key properties of the system, e.g., the value of $T_c$, the nesting characteristics of the electronic bands, as well as the nature of superconductivity. Our results provide critical information on the evolution of FS with temperature and are important for a microscopic understanding of high-temperature superconductivity and the associated phenomena in FeSe ML. Moreover, the results may provide guidelines for a possible tuning of $T_c$ in Fe-chalcogenide monolayers on oxide surfaces. In addition to the important observations discussed above, our work showcases how the dynamic charge response measured by means of high-resolution electron scattering techniques, e.g., HREELS can be used to probe any change in the size of FS. The idea can be implemented to many other quantum materials.

\section*{Acknowledgements}
We thank Markus D\"ottling  and Albrecht von Faber for technical assistance during some of the experiments. Kh.Z. acknowledges funding from the Deutsche Forschungsgemeinschaft (DFG) through the DFG Grants No. ZA~902/5-1 and No. ZA~902/8-1 and also the Heisenberg Programme ZA~902/3-1 and ZA~902/6-1.  Kh.Z. thanks the Physikalisches Institut for hosting the group and providing the necessary infrastructure. R.R. and K.Z. acknowledge the support of the Natural Sciences and Engineering Research Council of Canada (NSERC).

\bibliography {./Refs}

\end{document}